\documentstyle[12pt,epsfig]{article}

\begin{document}

\begin{center}
{\Large\bf The phase of the $\sigma \to \pi \pi$ amplitude in
$J/\Psi \to \omega \pi ^+ \pi ^-$}

\vskip 5mm
{D.V.~Bugg},   \\
{Queen Mary, University of London, London E1\,4NS, UK}
\end {center}

\begin{abstract}
The phase variation of the $\sigma \to \pi \pi$ amplitude  is
accurately determined as a function of mass from BES II data for
$J/\Psi \to \omega \pi ^+\pi ^-$.
The determination arises from interference with the
strong $b_1(1235)\pi$ amplitude.
The observed phase variation agrees within errors with that
in $\pi \pi$ elastic scattering.
\end{abstract}

PACS Categories: 11.80.Et, 13.20.Fe, 13.20.He, 14.40.Lb

\vskip 9mm

The $\sigma$ pole appears as a conspicuous $\pi ^+\pi ^-$ peak
in BES II data for $J/\Psi \to \omega \pi ^+\pi ^-$ [1].
This peak is absent from data on $\pi \pi$ S-wave elastic scattering.
The connection between these two processes is a question
which is explored here.

For both processes, the partial wave amplitude $f(s)$ may be written
\begin {equation}
f(s) = N(s)/D(s),
\end {equation}
where $N(s)$ has only left-hand cuts and $D(s)$ has only right-hand
cuts.
The $N$ function can be different for the two processes.
We pursue the hypothesis that $N(s)$ for $\pi \pi$ elastic
scattering contains an Adler zero, which is absent from
the production process.
The phase variation above the $\pi \pi$ threshold arises from
the right-hand cut.
The $D$ function should be the same for all processes if only a single
resonance contributes.
The question is whether BES data and $\pi\pi$ elastic scattering data
are consistent with this hypothesis.

\begin{figure} [t]
\begin{center}
\epsfig{file=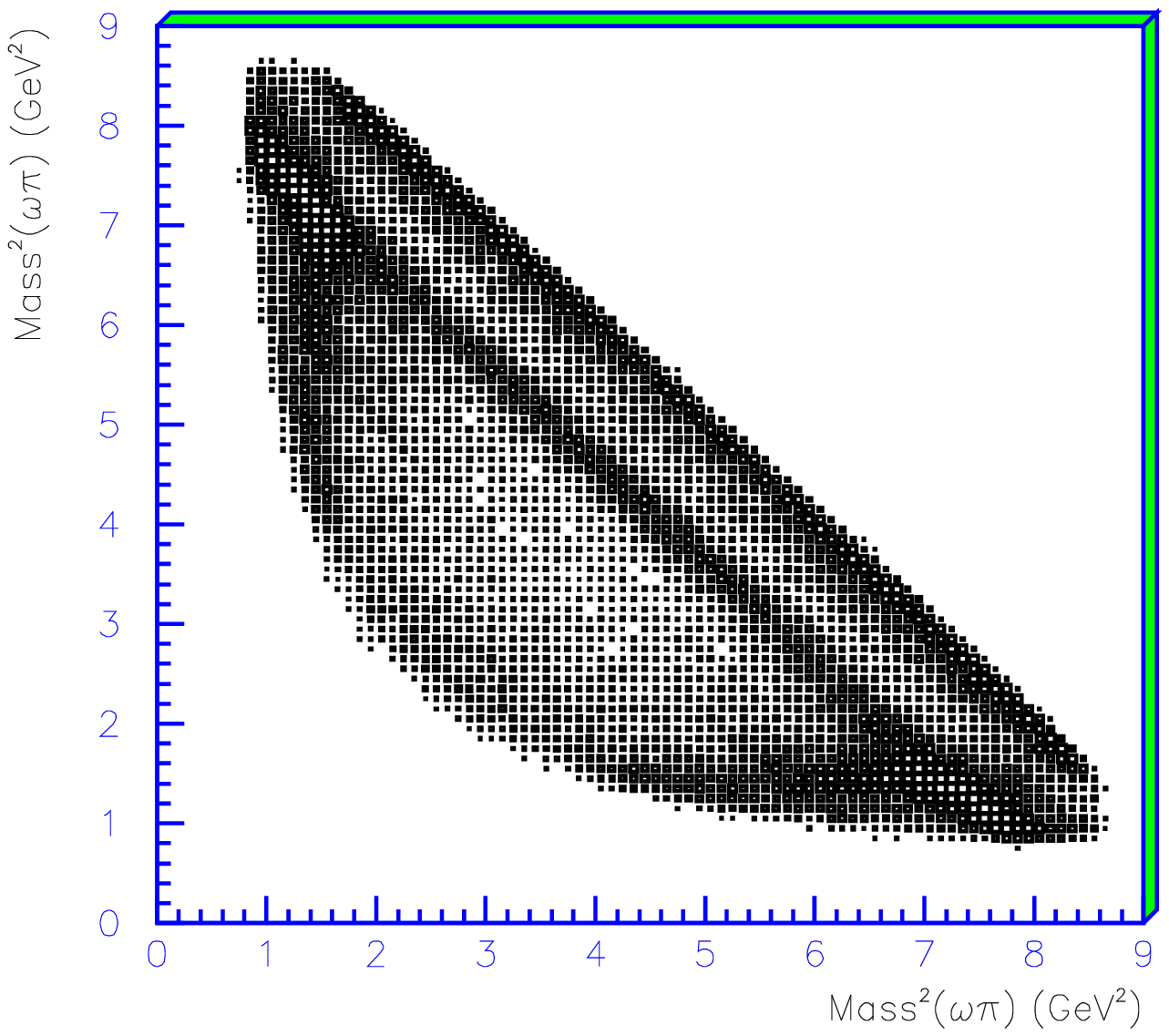,width=12.0cm}\
\caption[]{ The Dalitz plot for $\omega \pi ^+\pi ^-$.}
\end{center}
\end{figure}

The Dalitz plot for $J/\Psi \to \omega \pi ^+\pi ^-$ is shown in
Fig. 1.
The $\sigma$ pole appears as a diagonal band at the upper right-hand
edge of this plot.
There are also strong vertical and horizontal bands due to
$b^\pm _1(1235) \to \omega \pi ^\pm$.
These two bands account for 41\% of the data; the $\sigma$ pole
accounts for 19\% and $f_2(1270)$ for most of the remaining
intensity.
There is strong interference between the $b_1(1235)$ bands and the
$\sigma$ amplitude; this interference provides an accurate
determination of the phase $\delta _\sigma$ of the $\sigma$ as a
function of $\pi \pi$ mass.
The polarisation of the $\omega$ is along the normal to its decay
plane.
The $f_2(1270)$ components in the data have angular correlations
with this normal which are distinctively different from those of the
$\sigma$; as a result, $f_2$ and $\sigma$ are well separated in the
mass range where the $\sigma$ amplitude is sizable, up to 1000 MeV.
Above this mass, the $\sigma$ amplitude is swamped by the
$f_2(1270)$ peak.

The amplitude analysis follows the conventional isobar model.
The amplitude for the $b_1(1235) \pi$ final state is parametrised as
$\exp (i\Delta _{b1})F(b_1)$ and that for the $\omega$ is
parametrised as  $\exp (i\Delta _\sigma )F(\sigma \to \pi \pi )$.
Here $\Delta _{b1}$ and $\Delta _\sigma$ are constants describing the
strong interaction phases of the 3-body final states $b_1\pi$ and
$\omega \sigma$.
The $F(b_1)$ amplitude is a Breit-Wigner amplitude of constant
width for $b_1(1235)$.
A detail is that both S and D-wave decays of $b_1 \to \omega \pi$
are included, and the D/S ratio of amplitudes is fixed to the
PDG value of 0.29 [5].

The $F(\sigma \to \pi \pi )$ amplitude is taken as [3]:
\begin {eqnarray}
F(\sigma \to \pi \pi ) & = & \frac {G_{\sigma}}
{M^2 - s  - iM\Gamma _{tot}(s)}, \\
\Gamma _{tot}(s) &=& g_1\frac {\rho _{\pi \pi }(s)}{\rho _{\pi \pi
}(M^2)} + g_2\frac {\rho _{4\pi } (s)} {\rho _{4\pi }(M^2)}, \\
g_1 &=& (b_1 + b_2s)\frac {s - m^2_\pi /2}{M^2 - m^2_\pi /2}\exp [-(s -
M^2)/a].
\end {eqnarray}
Here $\rho _{\pi \pi }$ is the usual $\pi \pi$
phase space $2k/\sqrt {s}$ and $k$ is the momentum in the $\pi \pi$
rest frame. This formalism includes the Adler zero explicitly into
$\Gamma (s)$; the exponential factor cuts off the width at large $s$.
This formula has been fitted  simultaneously to BES data [1],
CERN-Munich data [4] and the $K_{e4}$ data of Pisluk et al. [5]. Our
objective is to determine the phase
\begin {equation}
\delta _\sigma (s) = \tan ^{-1}\left( \frac {M\Gamma (s)}{M^2 - s}
\right).
\end {equation}
A small detail is that eqn. (2) should strictly contain
a dispersive correction to the real part of the amplitude.
However, over the mass range covered here, this correction
is very small because the phase rises almost linearly with
$s$. The term $b_1 + b_2s$ fitted to the data accomodates
this small correction.

Another technical detail is that there are actually two $J/\Psi \to
\omega \sigma$ amplitudes having orbital angular momenta $L = 0$ and 2
in the production process. These are both included in the fit, with
different coupling constants and different strong interaction phases
$\Delta _\sigma$. A centrifugal barrier for production with $L=2$ is
included, but has little effect since the momentum in the
$\omega \sigma$ final state is large.
Likewise, $L = 0$ and 2 are both possible for
$J/\Psi \to b_1(1235)\pi$;
in practice the $L = 2$ amplitude is small.

In the fitting procedure, all amplitudes except that for
$\omega \sigma$ are fitted to the whole data set.
In order to determine the phase variation of the $\sigma$ amplitude
with mass,
slices 100 MeV wide are examined from $M_{\pi
\pi } = 400$ to 1000 MeV.
Lower masses are not accessible because the $b_1$ band runs off the
corner of the Dalitz plot; as a reminder, $s_{\pi \pi }$ varies
linearly as one moves perpendicular to the $\sigma$ band, with the
result that low masses are compressed tightly towards the edge of the
Dalitz plot.

The determination of $\delta _\sigma $ has been done in four
ways with progressively increasing freedom in the fit, in order to
check for consistency.
Results are shown as points with errors in panels (a)--(d) of Fig. 2.
In the first (most restrictive) approach (a), only one bin of
$\pi \pi$ mass is examined at a time.
The $\sigma$ amplitude is fitted to the whole $\pi \pi$ mass
range, but allowing a perturbation to the phase $\delta _\sigma$ of the
Breit-Wigner amplitude in a single bin.
In the second approach (b), both magnitude and phase of
$F(\sigma \to \pi \pi )$ are set free in one bin at a time.
In Fig. 2(c), the phase is set free in all bins
simultaneously, but the magnitude is fitted to the whole mass
range in accordance with eqns. (2)--(4).
In Fig. 2(d), the magnitude and phase are fitted freely
in all bins simultaneously.
Coupling constants of all other amplitudes are re-optimised for every
fit.

\begin{figure}[t]
\begin{center}
\epsfig{file=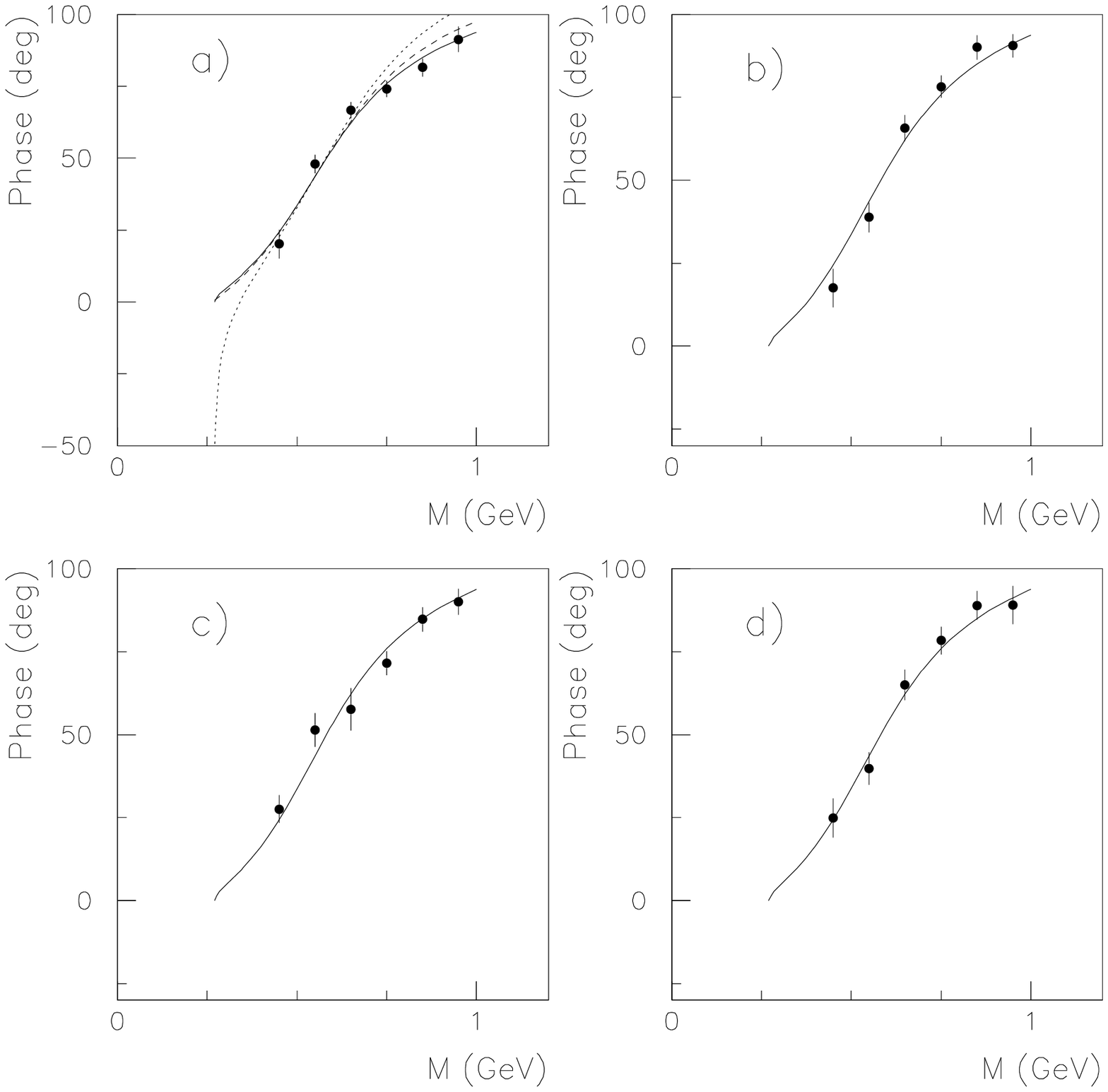,width=15.0cm}\
\caption{The phase of the
$\pi \pi$ S-wave amplitude.
The full curve shows the fit from eqns.
(2)--(4) to both BES data and elastic scattering data;
the dashed curve shows the fit with a Breit-Wigner amplitude of
constant width, and the dotted curve the fit with $\Gamma (s)
\propto \rho (s)$; points with errors show results fitted to slices of
$\pi \pi$ mass 100 MeV wide.
In (a), the phase is fitted one bin at a
time;
in (b) magnitude and phase are fitted one bin at a time;
in (c), phases are fitted in all bins simultaneously;
in (d) magnitudes and phases are fitted to all bins simultaneously.}
\end{center}
\end{figure}

The full curve of Fig. 2(a) shows the optimum fit to the whole mass
range using eqns. (2)--(4). The strong interaction phase
difference $\Delta _{b1} - \Delta _\sigma$ produces an offset,
which is furthermore different for $L = 0$ and $L = 2$ amplitudes;
only the phase variation with mass is meaningful. The
curve is therefore drawn so that $\delta _\sigma = 0$ at the $\pi \pi$
threshold. It turns out that the phases $\Delta _{b1}$ and $\Delta
_\sigma$ are such that the $\omega \sigma$ and $b_1\pi$ amplitudes
differ by $90^\circ$ in phase at a $\pi \pi$ mass of 600 MeV. The
interference term between the two amplitudes depends on the cosine of
the phase difference, and is therefore determined with maximum
sensitivity at this mass.

The dashed curve of Fig. 2(a) shows an alternative fit using for the
$\sigma$ a Breit-Wigner amplitude of constant width. In this case, the
offset $\Delta _{b1} - \Delta _\sigma$ is different
because the fitted mass $M$ is different; the curve is therefore
adjusted to reproduce the same phase as the full curve at 550 MeV, for
purposes of comparison. The dotted curve shows a fit using a
Breit-Wigner amplitude where $\Gamma (s) \propto  \rho _{\pi \pi }(s)$;
again it is fixed to the same phase as the full curve at 550 MeV,
to allow a clear comparison with the other two fits.

There is only small discrimination between the first two forms.
The agreement of phases with the curves demonstrates the correlation
of magnitude and phase expected from analyticity.
The third form, $\Gamma \propto \rho (s)$
(dotted curve), gives a somewhat poorer fit with slightly too large a
phase variation. It also suffers from the defect that it gives a
virtual state pole below the $\pi \pi$ threshold at $M_{\pi \pi } \sim
232$ MeV [3,6].

We consider the fit of Fig. 2(c) the most realistic.
In (b) and (d), there is statistical noise of
$\sim \pm 15\%$ in the intensity fitted to individual bins.
This noise is obviously unphysical, since the $\sigma$ amplitude
should vary smoothly with mass; noise in the magnitude introduces
noise into the phase via correlations in the real part of the
interference.
Errors on phases are therefore over-estimated in (b) and (d).

It comes as no surprise that $\delta _\sigma$ is accurately determined.
In Ref. [1], it was found that all three forms give pole positions
in agreement within $\pm 39$ MeV for the real part and
within $\pm 42$ MeV for the imaginary part.
The extrapolation from the Real $s$ axis to the pole requires that
real and imaginary parts of the $\sigma$ amplitude are separately
well determined.
This requires that the phase is also well determined as a function
of mass.

The determination of $\delta _\sigma$ is insensitive to the precise
mass and width of the $b_1$.
This is because the $\sigma$ and $b_1$ bands cross on the Dalitz plot
at an angle of 45$^\circ$ and the data integrate over the line-shape
of the $b_1$.
\begin{figure}[htbp]
\epsfig{file=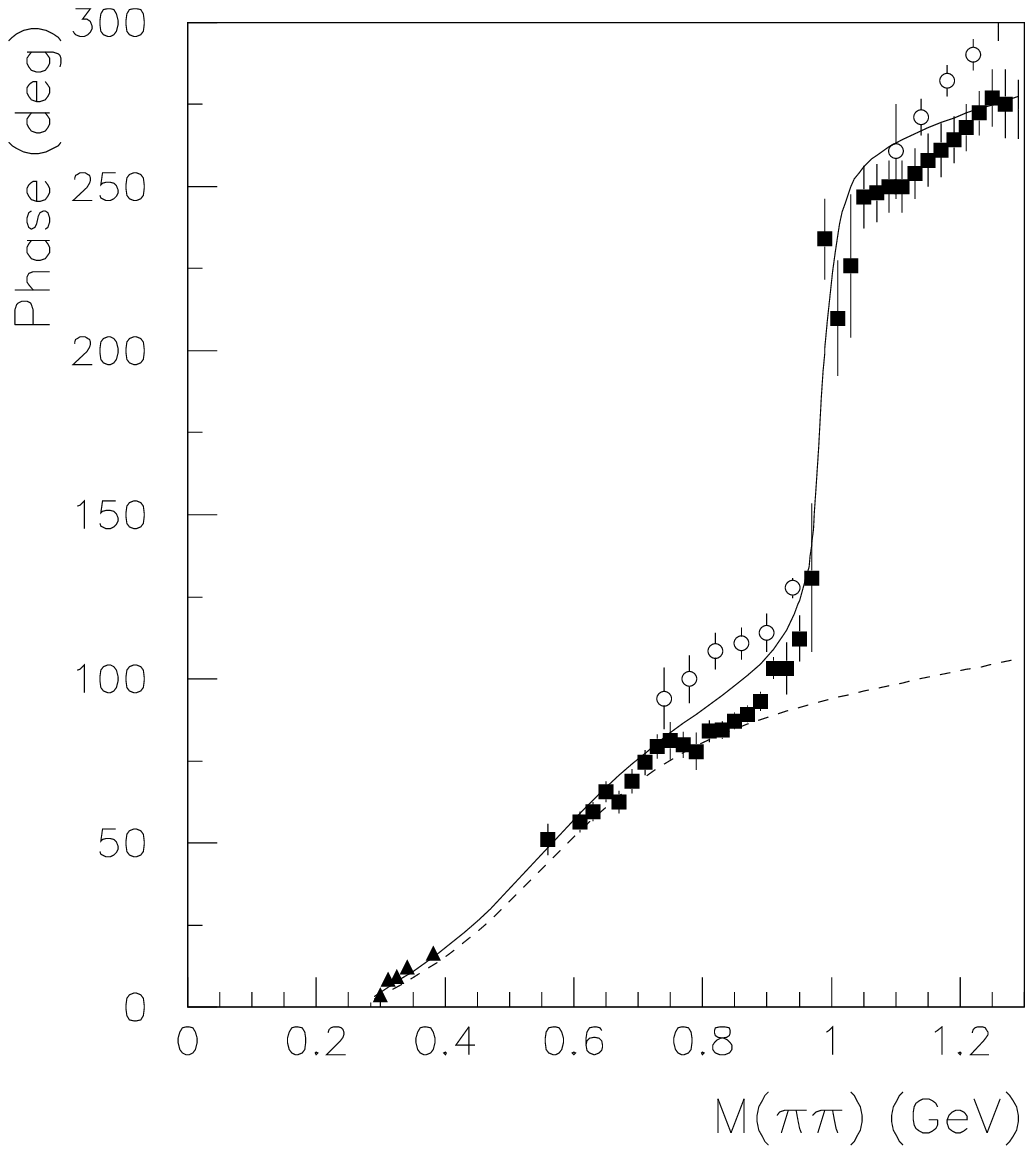,width=10cm}\
  \caption[]{Fit to elastic scattering data. Dashed curve: $\sigma$
  component from eqns. (2)--(4); full curve: full fit; triangles,
  $K_{e4}$ data [4]; black squares, Cern-Munich data [3];
  open circles, charge exchange data [8,9].}
\end{figure}

In fitting $\pi\pi$ elastic data, we adopt the Dalitz-Tuan prescription
[7], adding phases of $\sigma$ and $f_0(980)$ amplitudes. [The
$f_0(1370)$ and $f_0(1500)$ contributions are likewise added in, but
are very small]. This prescription guarantees that unitarity is
satisfied up to the inelastic threshold. The dashed curve of Fig. 3
shows the $\sigma$ phase $\delta _\sigma$ from eqn. (5); the full curve
shows the sum of all contributions. There is satisfactory agreement
with $K_{e4}$ data (triangles), CERN-Munich data (black squares) and
charge-exchange data (open circles), though there is some discrepancy
between the latter two above 700 MeV; the fit goes midway between these
two sets of data.

The phase information places restrictions on models of the $\sigma$.
Although a Breit-Wigner amplitude of constant width fits production
data, it gives the absurd result for elastic scattering that
$\delta _\sigma = 63^\circ$ at threshold.
This requires a compensating background phase of $\sim -63^\circ$ at
all masses; this is unphysical, since left-hand cuts cannot
reproduce such a behaviour.

A Breit-Wigner amplitude with $\Gamma \propto \rho (s)$ likewise
requires a large background phase in elastic scattering
$\sim -50^\circ$ at threshold.
A fit to elastic scattering then requires a background phase which
drops rapidly from zero at threshold to $\sim -50^\circ$.
The Ishida group has shown that elastic data may be fitted with a
repulsive background phase linearly proportional to centre of mass
momentum and a Breit-Wigner amplitude with $\Gamma \propto \rho (s)$;
the scattering length is rather larger than experiment [10].
A more complicated background phase corrects this defect [11]
and also remove the virtual-state pole below threshold.

Angular distributions are shown in Fig. 4 for four
ranges of $\pi \pi$ mass.
The angle $\chi$ is the azimuthal angle between the
production plane of $J/\Psi \to \omega X$ and
the decay plane $X \to \pi \pi$.
The angle $\theta _\omega$ is the production angle
of the $\omega$ in the $J/\Psi$ rest frame.
The angle $\alpha _\pi$ is the decay angle of the
$\pi ^+$ in the rest frame of $X$, taken with
respect to the direction of the recoil $\omega$.
The angle $\beta _\pi$ is the angle of the $\pi ^+$
with respect to the direction of $X$ in the rest
frame of the $\omega$.
The third distribution, $\cos \alpha _\pi$ departs
significantly from isotropy. This effect was observed
in the earlier DM2 data [12].
Up to $M(\pi \pi) = 700$ MeV, the departure from isotropy
is due entirely to interference with $b_1(1235)$;
above 800 MeV, interference with $f_2(1270)$ begins to
play a role.
Up to 800 MeV, there is no evidence for any significant
$\pi \pi$ D-wave amplitude.
In (d), one sees a strongly varying decay angular
distribution due to  $f_2(1270)$.
\begin{figure}[ht]
\epsfig{file=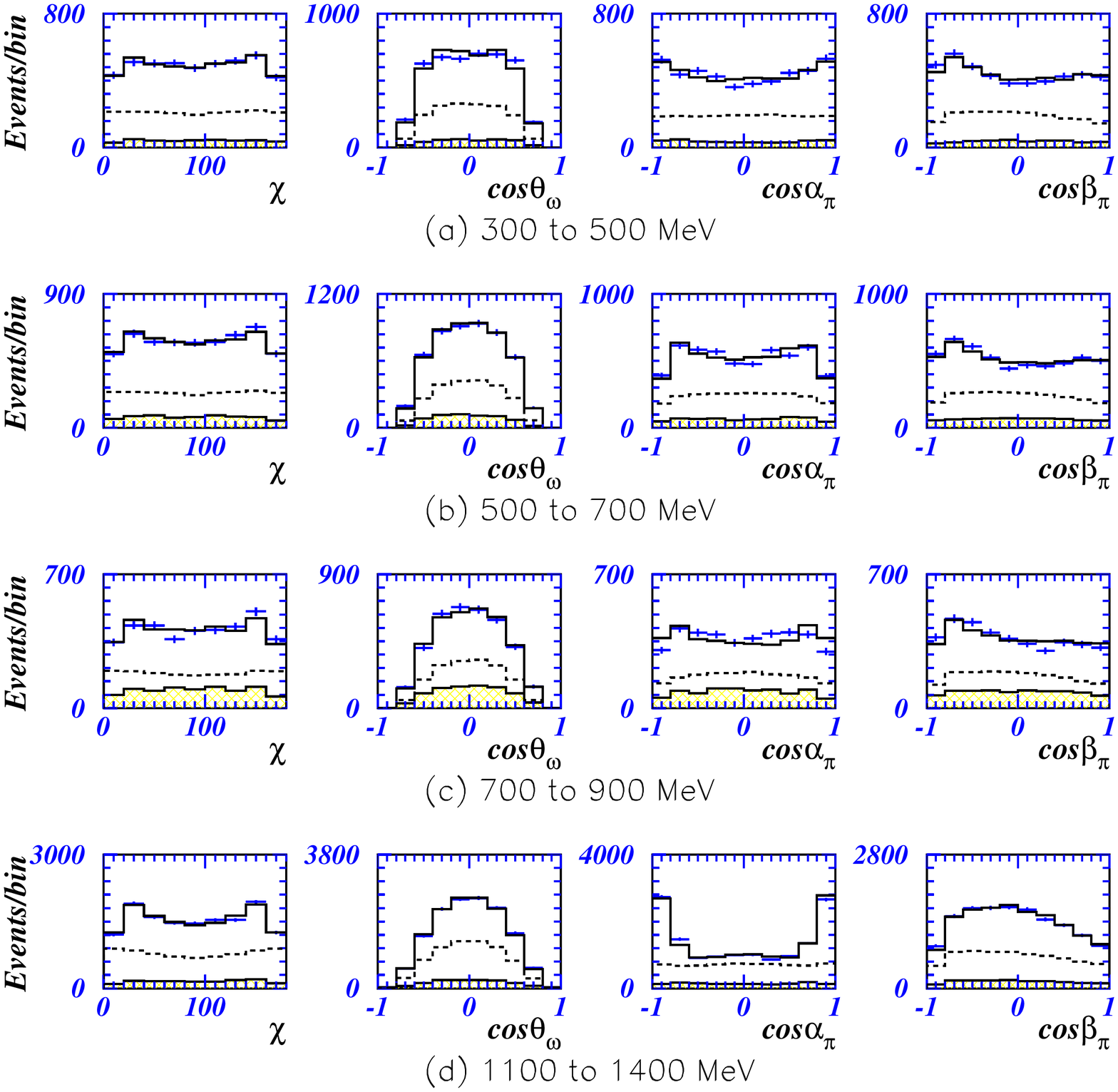,width=14cm}\
  \caption[]{Angular distributions (uncorrected for acceptance) for
  angles $\chi$, $\theta _\omega$, $\alpha _\pi$ and $\beta _\pi$
  defined in the text; histograms show the fit for four
  ranges of $\pi \pi$ mass. The lower histograms in each panel show
  backgrounds. Dashed curves show the acceptance.}
  \label{angl-wpp1}
\end{figure}

In summary, $\pi \pi$ elastic data and BES production data agree well
for the phase variation of the $\sigma$ amplitude with mass from
450 to 950 MeV.
This result is consistent with a single $\sigma$ resonance with
$s$-dependent width due to the Adler zero; however, some non-resonant
background phase could be present in addition.

I wish to thank the Royal Society for funding this work
and the BES collaboration for making the data available
as part of contracts Q772 between the Royal Society,
the Chinese Academy of Sciences and BES.

\end{document}